\documentclass[10pt,pra,aps,twocolumn,nofootinbib,preprintnumbers,superscriptaddress]{revtex4-2}

\usepackage{natbib}
\usepackage{amssymb,amsbsy,amsmath,amsfonts}
\usepackage{mathrsfs}
\usepackage{color}
\usepackage{graphicx}
\usepackage{slashed}
\usepackage{braket}
\usepackage[dvipsnames]{xcolor}
\usepackage{multibib}
\usepackage{mathtools}

\newcommand{\nn}{\nonumber}

\newcommand{\be}{\begin{equation}}
\newcommand{\ee}{\end{equation}}
\newcommand{\bea}{\begin{eqnarray}}
\newcommand{\eea}{\end{eqnarray}}
\newcommand{\chiTL}{\chi_{\rm TL}}

\newcommand{\vect}[1]{\mathbf{#1}}

\begin{document}

\title{Universal nondiffractive topological spin textures in vortex cores of light and sound}

\author{Elena Annenkova}
\affiliation{University of Bordeaux, CNRS, Laboratoire Ondes et Mati{\`e}re d'Aquitaine, F-33400 Talence, France}

\author{Andrei Afanasev}
\email{afanas@gwu.edu}
\affiliation{Department of Physics, The George Washington University, Washington, District of Columbia 20052, USA}

\author{Etienne Brasselet}
\email{etienne.brasselet@u-bordeaux.fr}
\affiliation{University of Bordeaux, CNRS, Laboratoire Ondes et Mati{\`e}re d'Aquitaine, F-33400 Talence, France}

\date{\today}

\begin{abstract}
We report universal skyrmionic spin textures in the cores of optical and acoustic vortex beams, described within the framework of Laguerre-Gaussian modes. We analytically demonstrate nondiffractive propagating spin merons, independent of whether the field is transverse or longitudinal, with their sign controlled by the wavefront helicity. Experimental confirmation is provided in acoustics through full three-dimensional measurements of the velocity vector field. Although these phenomena are intrinsic to vortex cores, we also show that the claimed universality breaks down for higher-order topological charges, depending on the carrier mode, here exemplified using the Bessel framework. 
\end{abstract}

\maketitle

{\it Introduction.}---Optical and acoustic waves are fundamentally different: the former are electromagnetic and can propagate in vacuum, while the latter are mechanical and require a material medium. They are also governed by distinct constitutive equations, giving rise to transverse or longitudinal fields. Despite these differences, light and sound share deep analogies, tracing back to ancient Greek theories attempting to unify the perceptions of seeing and hearing \cite{darrigol_centaurus_2010a, darrigol_centaurus_2010b}. Nowadays, such analogies are explored across a broad range of wave phenomena, including propagation effects such as refraction and reflection \cite{carcione_studia_2002}, waveguiding \cite{jen_ieee_1985}, and nonlinear responses \cite{bunkin_spu_1986}. Beyond propagation, analogies extend to mechanical action, through radiation forces on deformable interfaces \cite{bertin_prl_2012} or on particles manipulated by optical or acoustic tweezers \cite{thomas_jqsrt_2017}; see \cite{toftul_arxiv_2024} for a comprehensive review of optical and acoustic forces and torques. More recently, the concept of polarization and spin---long associated with light as the trajectory traced by the tip of the electric field vector---has been extended to sound waves \cite{shi_nsr_2019}. This framework enabled the identification \cite{bliokh_pof_2021} and observation \cite{muelas_prl_2022} of polarization singularities and topological polarization structures in sound, analogous to their optical counterparts \cite{bliokh_rpp_2019}. Even more recently, nondiffracting polarization structures around wavefield zeros have been predicted for both light \cite{afanasev_advphot_nexus_2023} and sound \cite{kille_prb_2024}, culminating in the observation of polarization-based skyrmions in the cores of optical vortex beams \cite{mata_prl_2025, mata_arxiv_2025}.

Here, we extend and unify these results by reporting the existence of universal, nondiffracting spin skyrmions in the cores of optical and acoustic vortex beams. Experimental demonstration is performed in the acoustic domain through three-dimensional (3D) measurement of the spin vector field and characterization of its topological structure. We further reveal the influence of the structure of the field carrying the singularity in the core region, which allows the generation of higher-order optical skyrmionic spin textures that rely on the spin-orbit coupling of light.

{\it General framework.}---To establish a common ground for both optics and acoustics, we consider monochromatic Laguerre-Gaussian (LG) beams with angular frequency $\omega$ (and further omit the time-dependent factors $e^{-i\omega t}$), which form a prototypical basis for vortex fields carrying an on-axis phase singularity of topological charge $l$. These beams are constructed from the solutions of the paraxial Helmholtz equation, whose expression in the cylindrical basis $(\vect{u_\rho}, \vect{u_\phi}, \vect{u}_z)$ reads
\bea
\nonumber
&f_{l,m}(\rho,\phi,z) \propto \frac{w_0}{w(z)} \Bigl(\frac{\rho}{w(z)}  \Bigr)^{\!|l|} \! \mathcal{L}^{l}_m\left( \! \frac{2\rho^2}{w(z)^2}\!\right) \! \exp \!\left (\!-\frac{\rho^2}{w(z)^2}\right)
\\
&\times \!\exp\left\{ i \left[l\phi + kz + k\frac{\rho^2}{2R(z)} - (|l|+2m+1) \psi(z) \!\right]\right\},
{\label{eq:LG}}
\eea
where $w(z) = w_0\sqrt{1+(z/z_0)^2}$ is the characteristic beam radius along the propagation axis, $k$ the wavenumber, $z_0 = kw_0^2/2$ the Rayleigh distance, $R(z) = z\bigl[1 + (z_0/z)^2\bigr]$ the radius of curvature, $\psi(z) = \arctan(z/z_0)$ the Gouy phase, and $\mathcal{L}^{l}_p$ the generalized Laguerre polynomial of azimuthal order $l \in \mathbb{Z}$ and radial order $m \in \mathbb{N}$.

We then introduce the reduced 3D spin vector $\vect{S}$ as
\be
\label{eq:Spin}
\vect{S} = \operatorname{Im}  \left[\vect{F}^* \times \vect{F} \right] / | \vect{F} |^{2} = |\vect{S}|\, \vect{s} \,,
\ee
where the unit spin vector $\vect{s}$ is introduced to characterize the topology of spin textures. It is defined from the local rotation of the underlying 3D vector field
 $\vect{F}(\vect{r})$, at the point $\vect{r}$ in space, according to the respective constitutive equations of light and sound. Specifically,
\be
\label{eq:F_sound}
\vect{F}_{\rm sound} (\vect{r})= \boldsymbol{\nabla} f_{l,m}(\vect{r}) / (i \varrho \omega)\,,
\ee
which corresponds to the velocity field of particles in a medium of density $\varrho$, and 
\be
\label{eq:F_light}
\vect{F}_{\rm light}(\vect{r}) = f_{l,m}(\vect{r})\,\vect{e}_\perp +\frac{i}{k} \boldsymbol{\nabla}_\perp \cdot \left[f_{l,m}(\vect{r})\,\vect{e}_\perp \right],
\ee
which correspond to the electric field, with $\vect{e}_\perp$ the 2D unit vector defining the polarization state in the transverse plane. This expression includes first-order nonparaxial corrections obtained from $\boldsymbol{\nabla} \cdot \vect{E} =0$ \cite{lax_pra_1975}. In the following, we restrict ourselves to the fundamental radial mode $m=0$ (so that $\mathcal{L}^{l}_0 \equiv 1$) and omit the corresponding indices. For the optical case, we consider circular polarization states, $\vect{e}_\perp = (\vect{x} + i\sigma\vect{y})/\sqrt{2}$, with $\sigma = \pm1$.

\begin{figure}[b!]
\centering
\includegraphics[width=1\columnwidth]{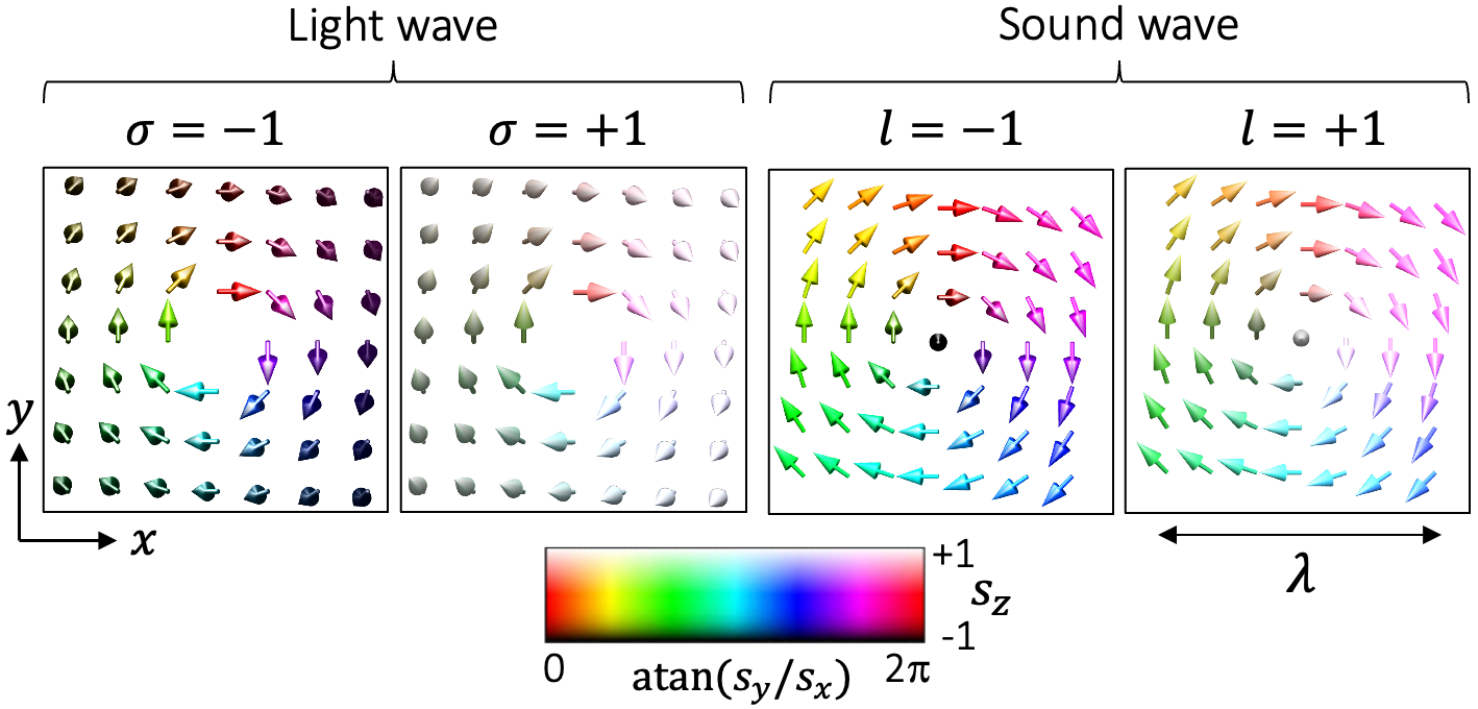}
\caption{Analytical core spin textures $\{{\vect{\widetilde s}}\}$ for unit-charge vortices of light (for $\sigma l<0$) and sound, over a $\lambda^2$ area centered on the $z$ axis from Eqs.~(\ref{eq:s_sound_core}) and (\ref{eq:s_light_core}). See Supplemental Materials, Sec.~II, for an alternative visualization emphasizing the sphere-to-plane stereographic projection of the spin texture.}
\end{figure}

{\it Nondiffracting core spin skyrmions.}---From Eqs.~(\ref{eq:LG}--\ref{eq:F_light}), the spin vector field $\vect{S}$ can be evaluated for arbitrary topological charge $l$ and position $\vect{r}$ in both optics and acoustics (see Supplemental Material, Sec.~I). Remarkably, deep inside the vortex core the unit spin vector field acquires a simple analytical form exhibiting well-defined topological textures. Introducing the dimensionless parameter $\xi \equiv (k\rho)^{-1}$ and considering the near-axis regime $\rho^2 \ll w^2$ , one obtains $\vect{s} \approx \vect{\widetilde s}$, with
\be
\label{eq:s_sound_core}
\vect{\widetilde s}_{\rm sound}= \frac{1}{\sqrt{1+l^2\xi^2}}(0,-1,l\xi)\ ,
\ee
and
\be
\label{eq:s_light_core}
\vect{\widetilde s}_{\rm light} = 
\begin{cases}
\frac{1}{\sqrt{1+4l^2\xi^2}}(0,-2|l|\xi,\sigma)\,, &  \sigma l < 0\,, \\
(0,0,\sigma)\,, & \sigma l > 0\,.
\end{cases}
\ee 

Propagation-invariant nontrivial topological textures confined within a wavelength-scale region near the singularity are thus obtained both for light and sound, provided $\sigma \,\ell < 0$ in the optical case, which points to the role of spin-orbit coupling in the emergence of nontrivial topologies in optics. This is illustrated in Fig.~1, where the mapping of the spin texture in the transverse plane onto the unit sphere exhibits the characteristic pattern of Bloch-type merons with Skyrme number $N=\pm 1/2$ for all $l$, defined as
\be
\label{eq:N}
N[\,\vect{\widetilde s}\,] = \frac{1}{4\pi} \iint \vect{\widetilde s} \cdot \left( \frac{\partial\,\vect{\widetilde s}}{\partial x}\times\frac{\partial\,\vect{\widetilde s}}{\partial y}\right) \rho d\rho d\phi\,.
\ee
Combining Eqs.(\ref{eq:s_sound_core}), (\ref{eq:s_light_core}) and (\ref{eq:N}), one obtains for the normalized spin textures  $N[\,\vect{\widetilde s_{\rm sound}}] = {\rm sign}(l)/2$ in acoustics, and $N[\,\vect{\widetilde s_{\rm light}}] = -\sigma/2 = {\rm sign}(l)/2$ in optics, consistent with the spin-orbit helicity condition $\sigma l < 0$ that defines the nontrivial meron topology. Also, we note that transverse-to-longitudinal versus longitudinal-to-transverse radial evolution of the spin vector directly reflects the underlying divergence-free (transverse) versus curl-free (longitudinal) character imposed by their respective constitutive equations.

\begin{figure}[t!]
\centering
\includegraphics[width=1\columnwidth]{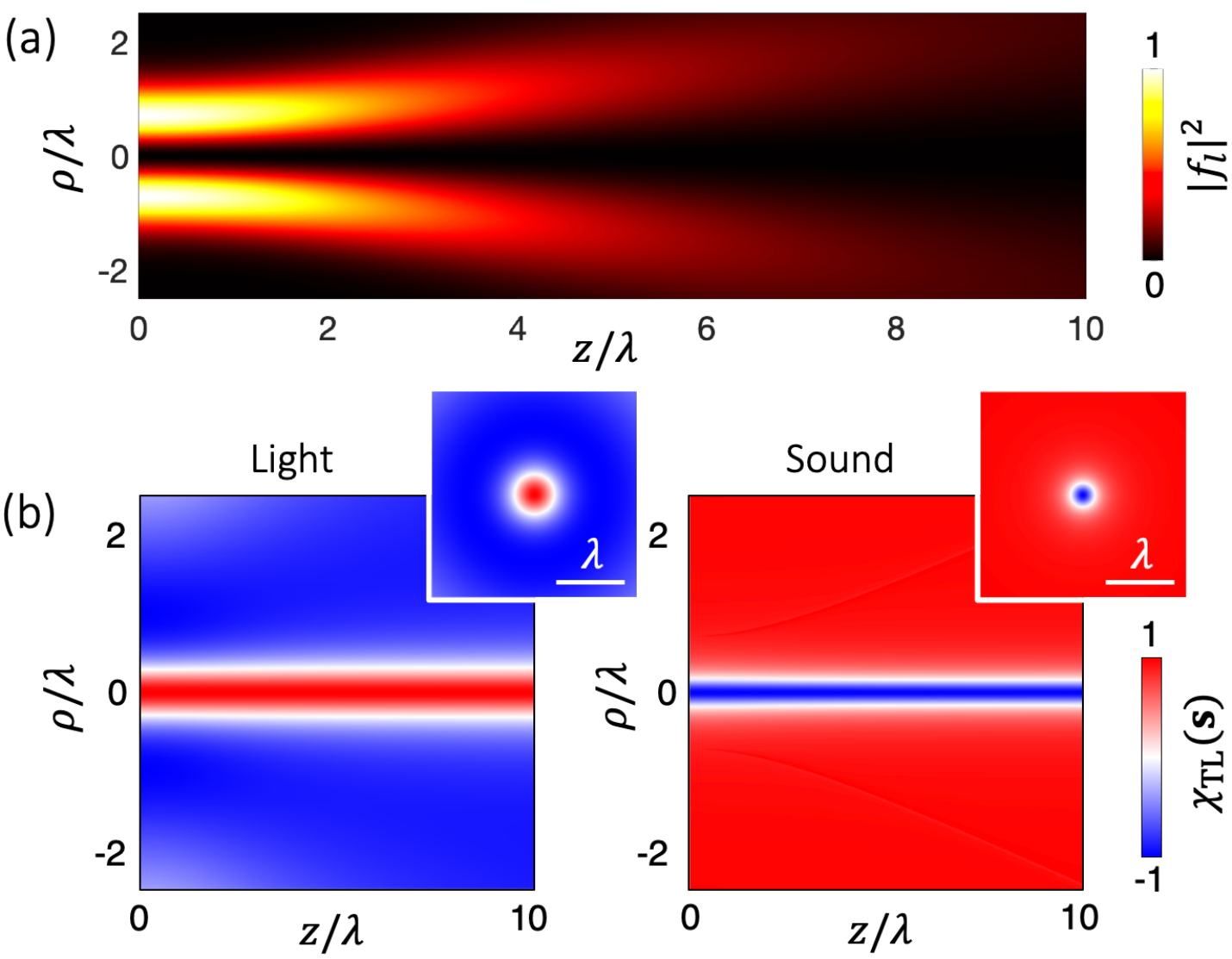}
\caption{(a) Diffracting intensity pattern $|f_l(\mathbf{r})|^2$ of a vortex beam with $l=\pm1$ whatever the nature of the wave, from Eq.(\ref{eq:LG}). (b) Calculated maps of $\chiTL(\vect{s})$ in the meridional plane $(z,\rho)$ and in the transverse plane at $z=0$ (insets), for light and sound, for $|l|=1$ with $\sigma l < 0$ for optics. All plots calculated for $w_0=\lambda$.}
\end{figure}

As suggested by Fig.~1, the topological charge of the spin skyrmion progressively builds up as the integration domain expands from the on-axis phase singularity. To quantify the localized character of these textures, we introduce the truncated Skyrme number $\bar N(\rho_0)$ by restricting the integration of Eq.~(\ref{eq:N}) to a disk of radius $\rho_0$ (see Supplemental Materials, Sec.~III). This allows extracting a characteristic radius $\rho_0^*$ enclosing a stable fractional skyrmion with ${\bar N} = \pm 1/4$, given by $\rho_0^* = \frac{\sqrt{3}}{2\pi} |l| \lambda$ for sound and $\rho_0^* = \frac{1}{\pi\sqrt{3}} |l| \lambda$ for light, thereby emphasizing the subwavelength confinement of the skyrmion. Its nondiffracting character is illustrated in Fig.~2(a), where the diffracting intensity profile of the vortex beams, $|f_l(\vect{r})|^2$, is contrasted with the propagation evolution of the transverse-to-longitudinal (TL) alignment parameter (see Supplemental Materials, Sec.~IV) of the spin, defined as
\be 
\label{eq:chi_TL}
\chiTL(\vect{s}) = |\vect{s}_{\perp}|^2 - s_z^2\,,
\ee 
which takes the values $\chiTL = +1$ for a purely transverse spin and $\chiTL = -1$ for a purely longitudinal spin, see Fig.~2(b) where $w_0=\lambda$. Furthermore, we note that the alignment parameter maps for the spin, $\chiTL(\vect{s})$ only marginally differs from its core analytical limit $\chiTL(\vect{\widetilde s})$ given by Eqs.~(\ref{eq:s_sound_core}) and (\ref{eq:s_light_core}) in the focal region of the beam ($|z| < z_0$), even in a markedly nonparaxial regime as illustrated in Fig.~2.

\begin{figure}[t!]
\centering
\includegraphics[width=1\columnwidth]{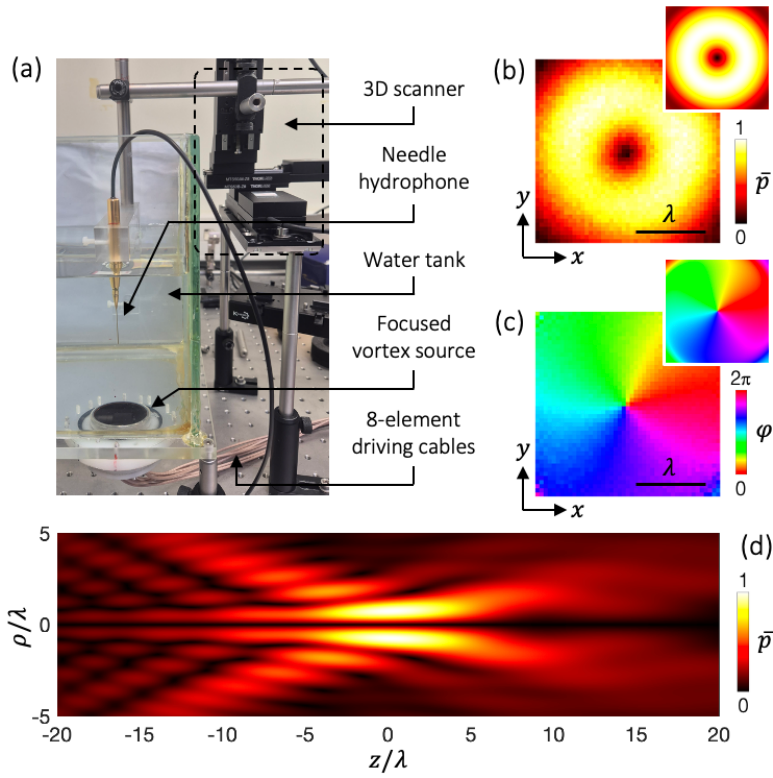}
\caption{(a) Photograph of the acoustic experimental setup. (b) Experimental normalized transverse pressure magnitude of a vortex beam with $l=1$, $\bar p=|p|/\max(|p|)$, at focus ($z=0$), over a $2.6 \lambda \times 2.6 \lambda$ area. Inset: simulation. (c) Corresponding pressure phase map, $\varphi = \arg(p)$. Inset: simulation. (d) Simulated meridional distribution of $\bar p$. Simulations are made using Eqs.~(\ref{eq:p_s}) and (\ref{eq:p}).}
\end{figure}

{\it Experimental framework.}---The observation of the predicted nondiffracting spin skyrmions are carried out in water, in the ultrasonic domain, by using a spherically shaped piezoelectric immersion transducer with a radius of curvature of 38.4~mm, radius $R = 19$~mm, central frequency $\nu = 2.25$~MHz, and a 600~kHz bandwidth. Its axisymmetric active surface is divided into eight impedance-matched sectors, each independently driven in parallel by a source emitting periodic sinusoidal bursts of 10 cycles at carrier frequency $\nu$ and a 100~kHz repetition rate, under a 70~V peak-to-peak voltage. Ultrasonic vortex beams are produced by adjusting the electrical phase delay applied to each sector according to the desired topological charge, while maintaining constant acoustic power per sector. The accessible range of topological charge is limited to $-3 \leq l \leq 3$ owing to the finite number of independent channels, see Supplemental Materials, Sec.~V, noting that imperfections from individual channels promote the splitting of high-charge phase singularities along the propagation axis. The acoustic field is 3D scanned with a spatial step of $50~\mu$m using either (i) a calibrated optical fiber hydrophone of 125~$\mu$m outer diameter with a 10~$\mu$m sensitive element, where $\lambda = c/\nu \simeq 0.66$~mm for a sound velocity $c = 1480$~m\,s$^{-1}$ in water, or (ii) a calibrated needle hydrophone with 200~$\mu$m diameter sensor, see Fig.~3(a).

The experimental generation of the acoustic vortex beam is illustrated in Figs.~3(b) and 3(c), which show the transverse spatial distributions of the pressure field magnitude, $|p|$, and phase, $\arg(p)$, for $l = 1$, which corresponds to $w_0 \simeq \lambda$. These observations are validated by simulating the generated acoustic field, assuming a complex source pressure distribution of the form
\be
\label{eq:p_s}
p_{\rm s}(r_{\rm s},\phi_{\rm s}) = p_0 \exp \left[ i l \phi_{\rm s} \!-\! ikF\!\left(1\!-\!\sqrt{1-(r_{\rm s}/F)^2}\right) \right]
\ee
where $p_0$ is a constant amplitude and $(r_{\rm s}, \phi_{\rm s})$ denote the polar coordinates of a point $\vect{r}_{\rm s}$ on the effective source plane. The total radiated pressure field is then obtained from the Rayleigh diffraction integral,
\be
\label{eq:p}
p(\vect{r}) = \frac{1}{i\lambda} \int_0^{2\pi}\!\!\!\int_0^R \!p_{\rm s}(\vect{r}_{\rm s}) \,\frac{\exp(ik|\vect{r}-\vect{r}_{\rm s}|)}{|\vect{r}-\vect{r}_{\rm s}|}\, r_{\rm s}dr_{\rm s}d\phi_{\rm s}\,,
\ee
with the focal plane defined at $z=0$. Although the generated acoustic vortex resembles an ideal Laguerre-Gaussian beam at focus, the associated field significantly departs from it upon propagation, as shown in Fig.~3(d), which reveals pronounced radial ringing that evolves with distance. Nevertheless, the phase singularity remains aligned with the $z$ axis throughout, ensuring that the wavefront topology remains equivalent to that of a Laguerre-Gaussian beam---a key feature governing the emergence of the spin texture in the vortex core---while expansion in terms of Bessel vortex modes is also possible \cite{wang2022electromagnetic}

\begin{figure}[b!]
\centering
\includegraphics[width=1\columnwidth]{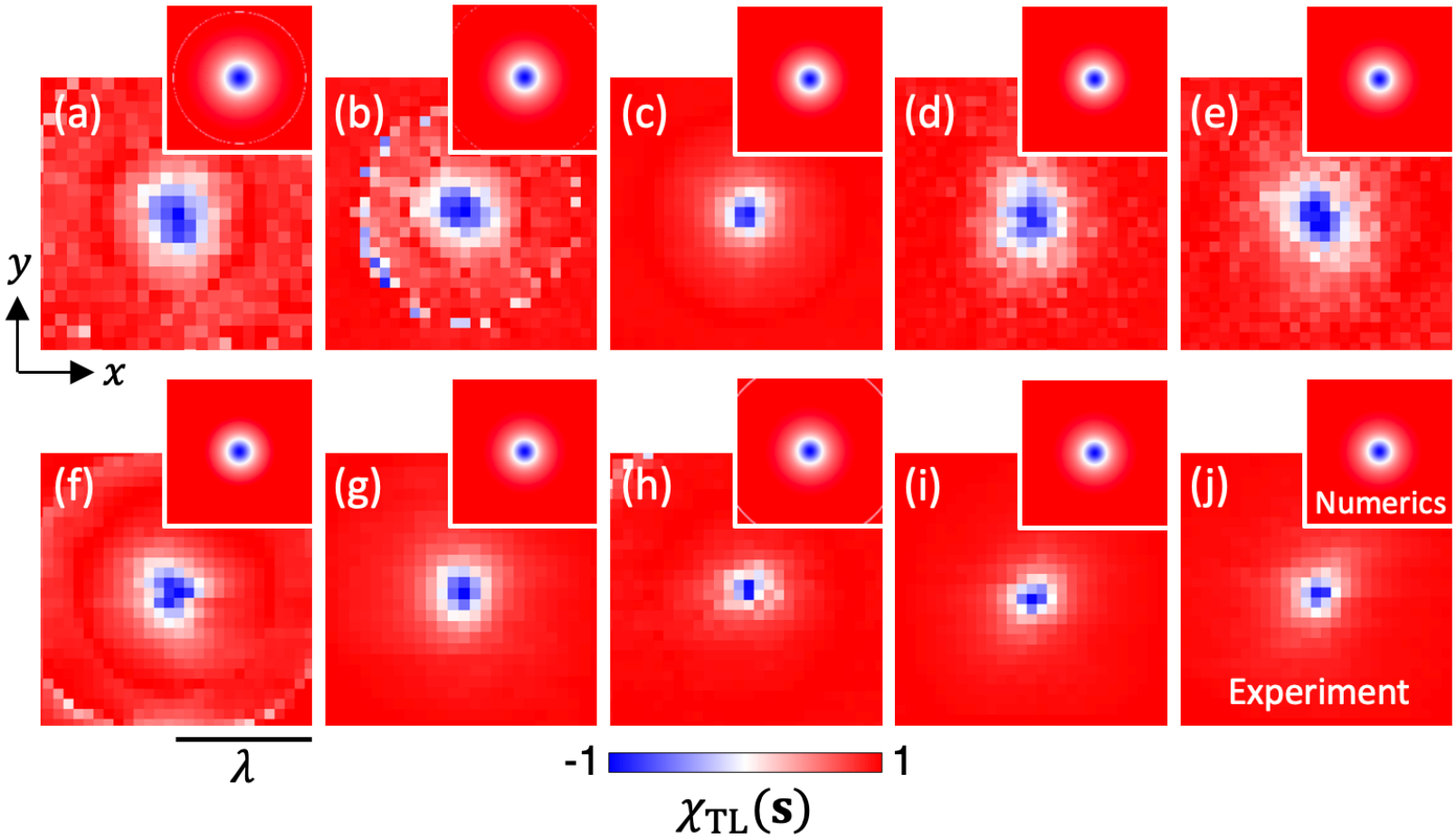}
\caption{Experimental transverse-to-longitudinal alignment parameter for the spin vector, $\chiTL(\vect{s})$, for an acoustic vortex beam with $l=1$ over a $4 \lambda^2$ area centered on the $z$ axis, in the range $-10~{\rm mm} < z < 35$~mm, with 5~mm increments. Insets: corresponding simulations from Eqs.~(\ref{eq:p_s}) and (\ref{eq:p}).}
\end{figure}

{\it Nondiffracting core spin textures.}---The full 3D spin field $\vect{s}$ is reconstructed from direct measurements of the complex pressure field, from which the vector velocity field is retrieved as $\vect{F}_{\rm sound}(\vect{r}) \propto \boldsymbol{\nabla} p(\vect{r})$. The nondiffracting character of the spin texture is evidenced in Fig.~4, which displays the evolution of $\chiTL(\vect{s})$ along the propagation direction over the range $z = -10$~mm to $z = 35$~mm ($-5 \lesssim z/z_0 \lesssim 17$). The observed quasi-tube-like evolution along $z$, well beyond the nominal focal region $|z| < z_0$, highlights the persistence and robustness of the core spin structure, even for vortex beams that significantly deviate from the ideal Laguerre-Gaussian mode.

\begin{figure}[t!]
\centering
\includegraphics[width=1\columnwidth]{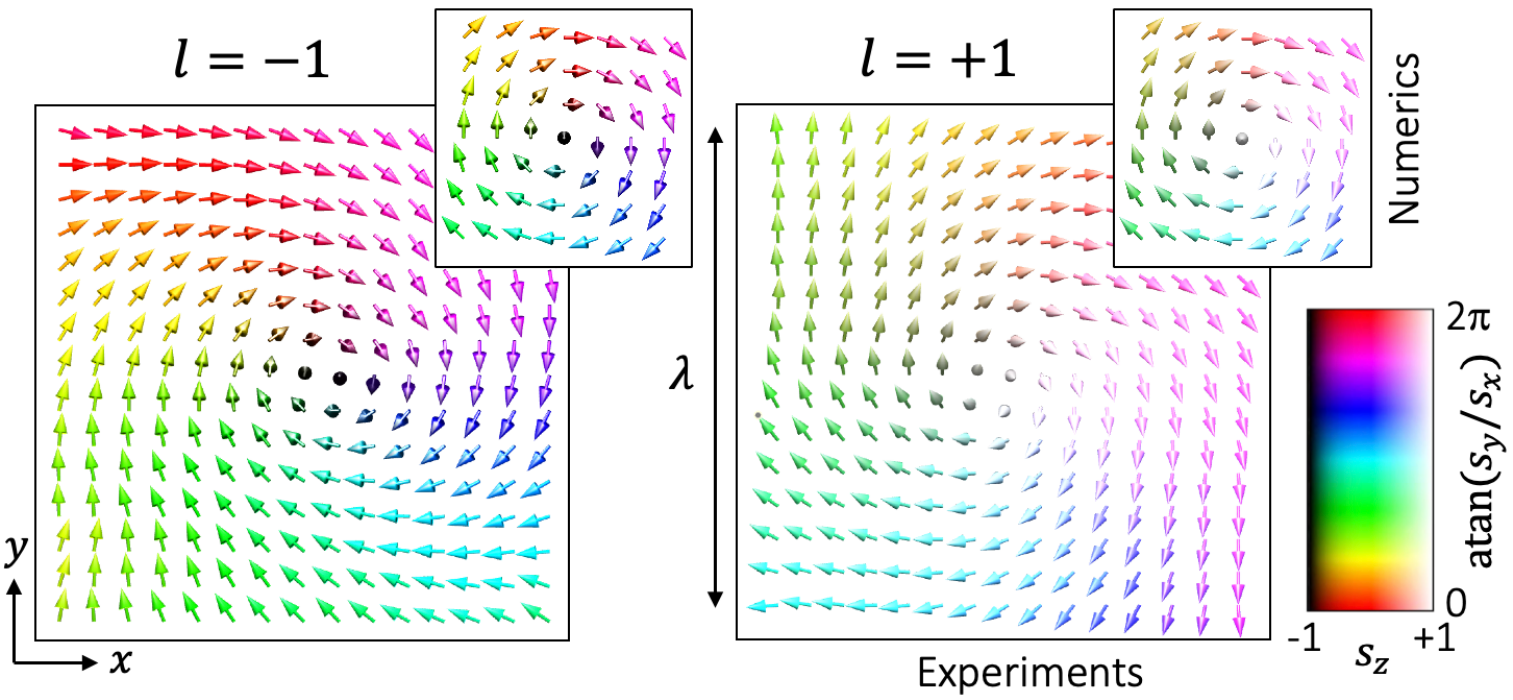}
\caption{Experimentally reconstructed textures $\{{\vect{s}}\}$ of spin core merons of the Bloch type for unit-charge  acoustic vortex beams with $l=\pm1$, over a $\lambda^2$ area centered on the $z$ axis, at $z=0$. Insets: corresponding simulations with half sampling density.}
\end{figure}

{\it Spin core merons for unit-charge vortices.}---The transverse acoustic spin textures reconstructed at the focal plane ($z=0$) for vortex topological charges $l = \pm 1$ are shown in Fig.~5. The spin vector field exhibits the characteristic Bloch-type meron structure: near the vortex center, the spin points mainly along $z < 0$ for $l = -1$ and along $z > 0$ for $l = +1$, while it continuously tilts into the transverse plane as the distance from the core increases, forming half-skyrmion patterns. These experimental observations confirm the analytical core prediction given by Eq.~(\ref{eq:s_sound_core}) and are in agreement with numerical simulations that incorporate the finite-size nature of the experimental transducer, with the acoustic field computed from the Rayleigh diffraction model [Eqs.~(\ref{eq:p_s}) and (\ref{eq:p})], as shown in the insets of Fig.~5. This confirms that the spin topology is correctly captured despite the experimentally generated field deviating from the ideal Laguerre-Gaussian beam due to the source geometry.

\begin{figure}[t!]
\centering
\includegraphics[width=1\columnwidth]{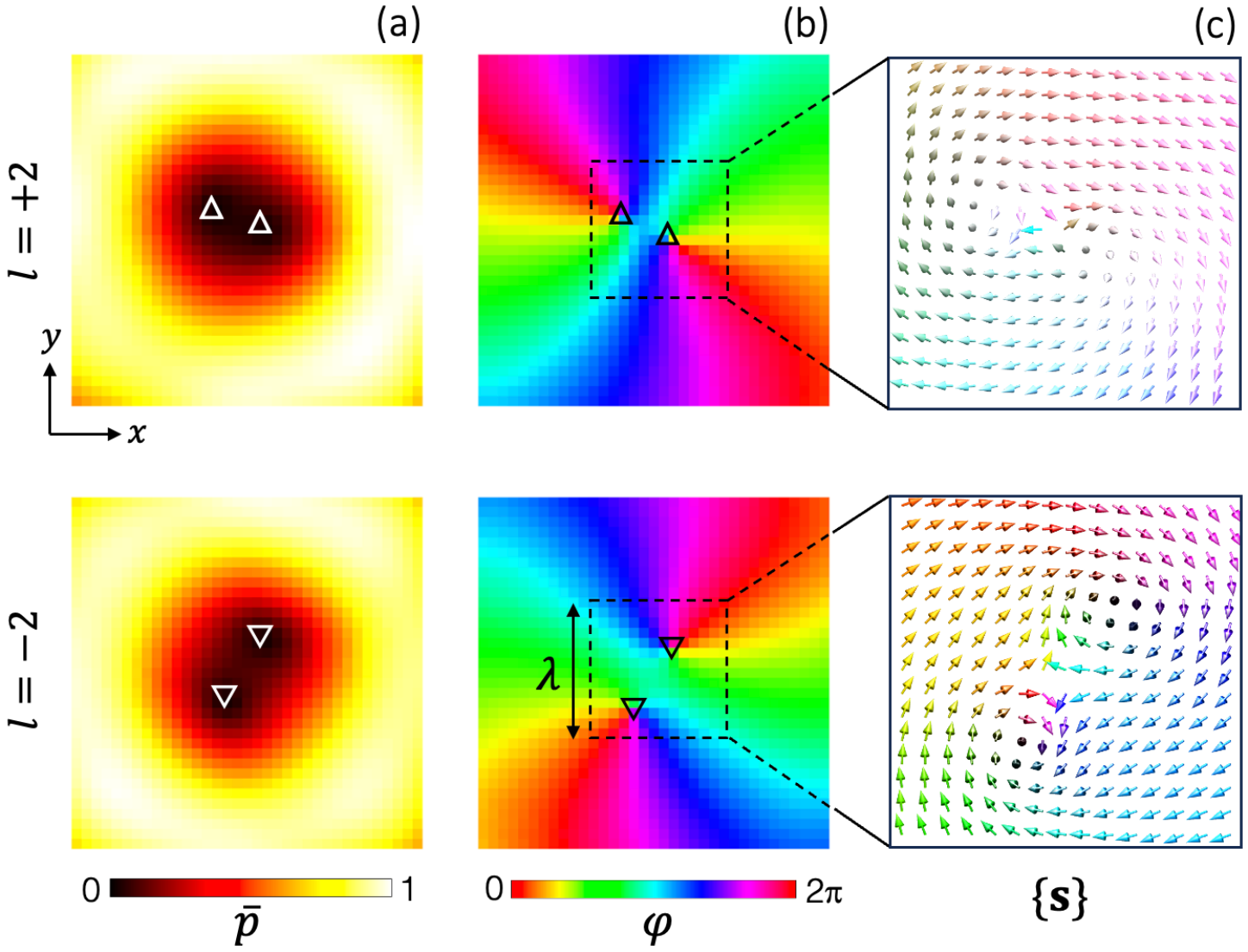}
\caption{Experimental core observation of high-charge acoustic vortex beams with $l = \pm 2$, over a $2.6 \lambda \times 2.6 \lambda$ area centered on the $z$ axis, at $z=0$. (a) Normalized transverse map of the pressure magnitude, $\bar p$. The triangle markers indicate the two unit-charge phase singularities ($\bigtriangleup$: positive charge; $\bigtriangledown$: negative charge) of the split charge-two vortex. (b) Corresponding pressure phase map, $\varphi$. (c) Enlargement of the triplet spin texture $\{ \vect{s}\}$ arising from high-charge vortex splitting, over a $\lambda^2$ area. Arrows color encoding is that of Fig.~5.}
\end{figure}

{\it Skyrmionic spin textures for $|l|>1$.}---The universal prediction of spin merons embedded in the cores of acoustic and optical vortex beams is independent of the topological charge $l$ of the vortex field. This has been tested for $l = \pm 2$ by adjusting the dephasing between adjacent sectors of the eight-element transducer, with results summarized in Fig.~6. As early noted in \cite{nye_prsla_1974}, high-order phase singularities are unstable in practical situations and inevitably split into $|l|$ unit-charge singularities, regardless the wave nature, as observed in Fig.~6(a) and 6(b). Such splitting transforms the ideal on-axis spin meron into a triplet spin texture shown in Fig.~6(c). 

{\it Beyond Laguerre-Gaussian framework.}---Recalling that the reported emergence of spin merons in the core of optical and acoustic vortex beams has been demonstrated in the paraxial Laguerre-Gaussian framework, we ask whether the core topology depends on the chosen vortex basis. We show that this is the case, as illustrated by considering optical Bessel vortex beams formed by superposition of plane waves, with helicity $\sigma$, whose wavevectors lie on a conical surface making an angle $\theta$ with respect to the propagation axis $z$ \cite{Durnin87, Jaregui05}. Within the paraxial approximation $\theta \ll 1$, we obtain from the Taylor expansion of Bessel functions, $J_n(x) \approx \frac{1}{n!}(\frac{x}{2})^n$, that the core field expressions are the same for Bessel and Laguerre-Gauss for $|l| = 1$, but differ for $|l| > 1$, which ultimately produces to distinct spin textures. In acoustics, one gets from Ref.~\citenum{bliokh_prb_2019}, in the core limit, 
\be
\label{eq:s_sound_core_Bessel}
\vect{\widetilde s\,}_{\rm sound}^{\rm Bessel}=  \vect{\widetilde s\,}_{\rm sound}^{\rm Laguerre{\text-}Gauss}\,,
\ee
which does not alter our previous conclusions. In optics, one gets (see Supplemental Materials, Sec.~VI)
\be
\label{eq:s_light_core_Bessel}
\vect{\widetilde s\,}_{\rm light}^{\rm Bessel} = 
\begin{cases}
\frac{[0,-2|l|\xi,\sigma(1-|l|(|l|-1)\xi^2)]}{\sqrt{1+2|l|(|l|+1)\xi^2+l^2(|l|-1)^2\xi^4}}
\,, &  \sigma l < 0\,, \\
(0,0,\sigma)\,, & \sigma l > 0\,.
\end{cases}
\ee
This gives a half-integer Skyrme number $N = -\sigma/2$ for $|l|=1$ an integer Skyrme number $N = -\sigma$ for $|l|>1$, provided that $\sigma l < 0$, which highlights how universality regarding the wave nature breaks due to spin-orbit interaction.

{\it Conclusion}---We have reported a universal form of spin topology emerging in the cores of optical and acoustic vortices. Our results reveal how singular wavefronts naturally host nondiffracting skyrmionic spin textures, independent of whether the field is transverse or longitudinal. We have further shown how this universality breaks depending on the modal basis due to optical spin-orbit coupling. Experimental demonstration made in the acoustic domain from full 3D field measurements could be extended to optics by using polarization-resolved far-field analysis of light scattered by a nanoparticle acting as a local probe \cite{eismann_acsphot_2024}, thereby enabling direct access to the longitudinal spin component without need to rely on Gauss's law as done in Ref.~\cite{mata_prl_2025}. Looking forward, the inherent spin structuring and nondiffracting character of these switchable topological textures may inspire new concepts in imaging and information transfer, while offering a versatile framework to study spin-orbit coupling and chiral light-matter interaction at the subwavelength scale. Moreover, three dimensional optical spin topologies may map into spin polarization and magnetization of matter \cite{afanasev2020polarization}, opening new opportunities to generate magnetic skyrmions from optical ones.



\section*{SUPPLEMENTAL MATERIALS}

\section{Spin of Laguerre-Gaussian beams}

\subsection{Acoustics}

From Eqs.~(1) and (3) of the main text, the velocity field $\vect{F}_{\rm sound}$ of a Laguerre-Gauss acoustic vortex beam defined of charge $l$ and radial index $m=0$ is given by
\be 
\vect{F}_{\rm sound} =F_{{\rm sound},\rho} \vect{u}_\rho + F_{{\rm sound},\phi}\vect{u}_\phi + F_{{\rm sound},z} \vect{u}_z
\ee 
 where
\bea
&&F_{{\rm sound},\rho}(\mathbf{r}) = \frac{f_l(\mathbf{r})}{i\varrho \omega}\Biggl(\frac{|l|}{\rho}+ \frac{ik\rho}{R(z)} - \frac{2\rho}{w(z)^2}\Biggr),\\
&&F_{{\rm sound},\phi}(\mathbf{r}) =\frac{f_l(\mathbf{r})}{i\varrho \omega}\Biggl(\frac{il}{\rho} \Biggr), \\
\nonumber
&&F_{{\rm sound},z}(\mathbf{r}) = \frac{f_l(\mathbf{r})}{i\varrho\omega}\Biggl\{-\frac{w_0^2}{z_0^2 w(z)^2}\Biggl[z+|l|z - \frac{2\rho^2z}{w(z)^2}\Biggr] \\
&&+\,ik\Biggl[1-\frac{\rho^2}{R(z)^2} + \frac{\rho^2}{2zR(z)} - \frac{w_0^2(|l|+1)}{kz_0 w(z)^2}\Biggr]\Biggl\}. 
\eea
Inserting above expressions in Eq.~(2) of the main text, one gets for cylindrical components of the spin vector, up to a common prefactor,
\bea
&&\hspace{-5mm} S_{{\rm sound},\rho} (\mathbf{r}) \propto |f_l(\mathbf{r})|^2 \frac{2 l w_0^2z(-2\rho^2+(1+|l|)w(z)^2}{\rho z_0^2 w(z)^4},\\
\nonumber
&&\hspace{-5mm} S_{{\rm sound},\phi}(\mathbf{r}) \propto |f_l(\mathbf{r})|^2\frac{1}{\rho w^4(z)}\Bigg\{ \Big(2\rho^2-|l|w(z)^2\Big) \\
\nonumber
&&\hspace{-5mm} \times\Bigg[ \frac{k\rho^2(z_0^2-z^2)w(z)^2}{z^2R(z)^2}-\frac{2}{k}\Big(2+2|l|-k^2w(z)^2 \Big)\\
&&\hspace{-5mm} +\, \frac{4 \rho^2 z}{z_0 R(z)}\Bigg]-\frac{4\rho^2z}{z_0 R(z)} \Bigg\}\,,\\
&&\hspace{-5mm} S_{{\rm sound},z}(\mathbf{r}) \propto |f_l(\mathbf{r})|^2  \Bigg(\frac{2|l|l}{\rho^2}-\frac{4l}{w(z)^2}\Bigg).
\eea

\subsection{Optics}

From Eqs.~(1) and (4) of the main text, the electric field $\vect{F}_{\rm light}$ of a Laguerre-Gauss acoustic vortex beam defined of charge $l$ and radial index $p=0$ is given by
\be 
\vect{F}_{\rm light} =F_{{\rm light},\rho} \vect{u}_\rho + F_{{\rm light},\phi}\vect{u}_\phi + F_{{\rm light},z} \vect{u}_z
\ee 
where
\bea
\label{eq:LGfields}
F_{{\rm light},\rho}(\mathbf{r}) &=& \frac{f_l(\mathbf{r})}{\sqrt{2}}\exp(i\sigma\phi),\\
F_{{\rm light},\phi}(\mathbf{r}) &=& \frac{i \sigma   f_l(\mathbf{r})}{\sqrt{2}} \exp(i\sigma\phi),\\
\nonumber
F_{{\rm light},z}(\mathbf{r}) &=& \frac{i f_l(\mathbf{r})}{\sqrt{2}} \frac{\rho w_0}{z_0 w(z)}\exp(i\sigma\phi)\\
& \times & \Bigg \{ \frac{|l|-\sigma l}{2\rho^2} w_0w(z)  - \exp[-i\psi(z)]\Bigg \}.
\eea
Inserting above expressions in Eq.~(2)of the main text, one gets for cylindrical components of the (electric) spin vector, up to a common prefactor,
\bea
&&S_{{\rm light},\rho} (\mathbf{r}) \propto \sigma |f_l(\mathbf{r})|^2 \frac{\rho}{R(z)},\\
&&S_{{\rm light},\phi}(\mathbf{r}) \propto -|f_l(\mathbf{r})|^2\Bigg( \frac{|l|-\sigma l}{k\rho} - \frac{2\rho}{kw(z)^2}\Bigg),\\
&&S_{{\rm light},z}(\mathbf{r}) \propto \sigma |f_l(\mathbf{r})|^2.
\eea

\begin{figure}[t!]
\centering
\includegraphics[width=1\columnwidth]{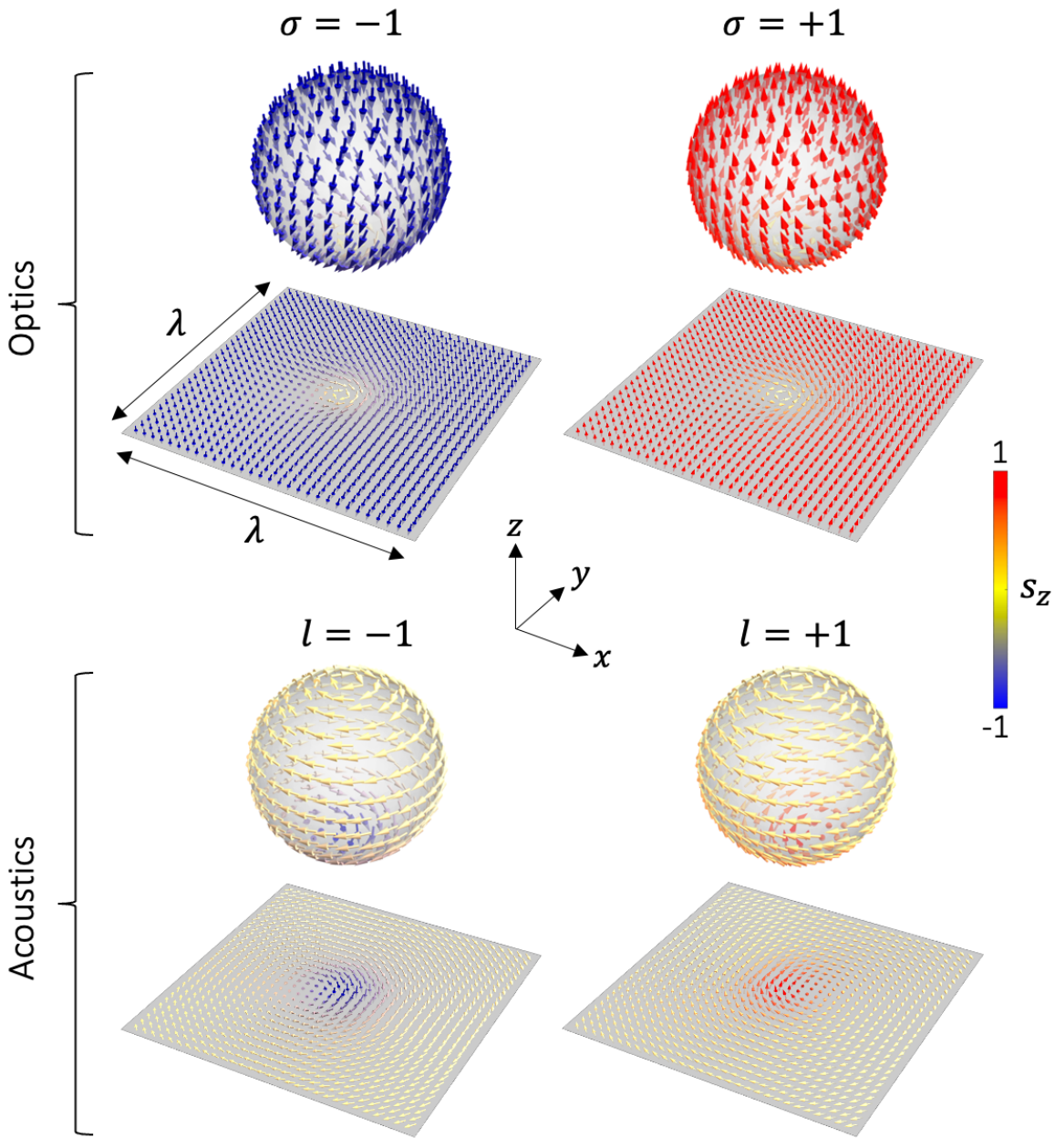}
\caption{Stereographic projections of optical and acoustic spin core merons for unit-charge vortex fields.}
\end{figure}

\section{Stereographic spin projections}

Figure~S1 illustrates the stereographic projections of the spin textures discussed in the main text. These plots provide an alternative visualization of the optical and acoustic spin merons embedded in the core of unit-charge vortex fields.

\section{Truncated Skyrme number}

In order to appreciate how the spin skyrmion in the core of a vortex beam is formed, we introduce the truncated Skyrme number $\bar N(\rho_0)$ by restricting the integration of the Skyrme density to a disk of radius $\rho_0$, namely
\be
\label{eq:N_truncated}
{\bar N} [\,\vect{\widetilde s}\,](\rho_0) = \frac{1}{4\pi} \! \int_0^{2\pi} \!\!\!\int_0^{\rho_0} \!\vect{\widetilde s} \,\cdot \left( \frac{\partial\, \vect{\widetilde s}}{\partial x}\times\frac{\partial \,\vect{\widetilde s}}{\partial y}\right)\rho d\rho d\phi\,.
\ee
By construction, $\bar N(\rho_0)$ characterizes the progressive buildup of the topological charge as the integration domain expands from the on-axis phase singularity outward. In the limit $\rho_0 \to \infty$, one recovers the full Skyrme number, namely $N[\,\vect{\widetilde s}_{\rm sound}]= {\rm sgn}(l)/2$ and $N[\,\vect{\widetilde s}_{\rm light}] = -\sigma/2$ as mentioned in the main text.

Inserting Eqs.~(5) and (6) of the main text in Eq.~(\ref{eq:N_truncated}), we obtain the following analytical expressions
\be
\label{eq:N_trunc_sound}
{\bar N}[\,\vect{\widetilde s}_{\rm sound}\,](\rho_0) \!=\! \frac{{\rm sgn}(l)}{2} \!\!\left[ 1 - \frac{1}{\sqrt{1+(k\rho_0/l)^2}} \right]\!,
\ee
and
\be
\label{eq:N_trunc_light}
{\bar N}[\,\vect{\widetilde s}_{\rm light}\,](\rho_0) = - \frac{\sigma}{2} \frac{k\rho_0/|l|}{\sqrt{4+(k\rho_0/l)^2}}\,,
\ee
 which allows defining the radius $\rho_0^*$ of the disk hosting a fractional skyrmion with ${\bar N}^*=\pm1/4$. We find $\rho_0^* = \frac{\sqrt{3}}{2\pi} |l| \lambda \simeq 0.28 |l| \lambda $ for sound and $\rho_0^* = \frac{1}{\pi\sqrt{3}} |l| \lambda \simeq 0.18 \lambda |l|$ for light.

\section{Transverse-to-longitudinal alignment parameter}

Whereas optical (acoustic) monochromatic plane waves are purely transverse (longitudinal) with respect to  the propagation direction, in structured monochromatic constituted of a discrete or continuum of plane waves, the electric (velocity) field has both longitudinal and transverse components. A concise description of the relative transverse-to-longitudinal (TL) field distributions in space for an arbitrary vector field $\vect{A}$ is provided by defining a TL-alignment parameter with respect to an arbitrary direction given the unit vector $\vect{n}$ as
\be 
\label{eq:chi_LT_general}
\chiTL(\mathbf{A}) = \frac{ |\vect{A}_{\perp}|^2 - |\vect{A}_{\parallel}|^2 }{ | \vect{A}|^2 },
\ee
where $\vect{A}_{\parallel} = (\vect{A} \cdot \vect{n})\vect{n}$ and $\vect{A}_{\perp} = \vect{A} - \vect{A}_{\parallel} $ are the longitudinal and transverse components of the vector field with respect to $\vect{n}$, satisfying $ | \vect{A}|^2=|\vect{A}_{\perp}|^2 + |\vect{A}_{\parallel}|^2$. In the main text, the director $\vect{n}$ is chosen as the beam propagation direction, $\vect{u}_z$, and we discuss the TL-alignment for both the exact spin vector field of Laguerre-Gauss vortex beams ($\vect{s}$) and its core  limit ($\,\vect{\widetilde s}\,$), whose analytical expression can be obtained from Eqs.~(5) and (6) of the main text, namely
\be
\label{eq:chiTL_sound_core}
\chiTL(\vect{\widetilde s}_{\rm sound}) = \frac{1-l^2\xi^2}{1+l^2\xi^2}
\ee
and
\be
\label{eq:chiTL_light_core}
\chiTL(\vect{\widetilde s}_{\rm light}) = 
\frac{4l^2\xi^2-1 }{4l^2\xi^2+1}
\ee

\section{Analytic Formalism for Vortex Topologies from Discrete Channels}

In order to evaluate the range of parameters required to observe a nondiffractive acoustic spin skyrmionic textures, we consider a simplified model of our experimental setup that admits an analytical solution. We thus consider a circular phased array of radius $R$, lying in the transverse plane and centered on the $z$ axis, composed of $M$ point-like acoustic sources, with a phase difference of $2\pi n/M$ between adjacent elements, resulting in a total phase shift of $2\pi n$ around the array. For array radii much larger than the wavelength, $R\gg \lambda$, the complex pressure field can be conveniently expanded in cylindrical coordinates using a Fourier-Bessel series \cite{wang2022electromagnetic},
\begin{align}
\label{eq:FBessel}
     &p_{\rm 0D}(\mathbf{r})=  p_0 \,
     \frac{e^{ik\sqrt{z^2+R^2}}
     }{\sqrt{z^2+R^2}}\,\times \nn \\ 
     &\sum_{m \in {\mathbb Z}} \left[ (-i)^l J_l(k_\perp \rho)e^{il(\phi-\phi_0)}\right]_{l=n+mM},
\end{align}
where 
the subscript 0D refers to the point-like nature of the individual sources, $k_\perp=kR/\sqrt{z^2+R^2}$, $p_0$ is a constant amplitude, $\phi_0$ is a reference angle ($i.e.$, an angular position of $M$th element of the array). Extension of Eq.~(\ref{eq:FBessel}) to the case of $N$ back-to-back arcs is obtained by superimposing an infinite number of $M$-point arrays rotated by a range of angles $\phi$ up to $\pm \pi/M$. It corresponds to making a reference angle $\phi_0$ an integration variable and averaging within a range $\phi_0=[-\pi/M,\pi/M]$, yielding a sinc-factor (where sinc$(x)\equiv \frac{\sin(x)}{x})$ for each term in the sum appearing in Eq.~(\ref{eq:FBessel}) that now reads as
\begin{align}
\label{eq:SecBessel}
&p_{\rm 1D}(\mathbf{r},t)=  p_0 \,
     \frac{e^{ik\sqrt{z^2+R^2}}
     }{\sqrt{z^2+R^2}}\,\times \nn \\ 
&\sum_{m \in {\mathbb Z}} \left[ (-i)^l J_l(k_\perp \rho)e^{il\phi}\mathrm{sinc}\left({\frac{\pi l}{M}}\right)\right]_{l=n+mM},
\end{align}
where the subscript 1D refers to the line nature of the individual sources. For an eight-channel ring-shaped actuator ($M=8$) aiming at producing a charge one vortex field, the applied phase parameter is $n=1$, and the lowest-order Bessel terms in the sum Eq.~(\ref{eq:SecBessel}) are $l=1,-7$ and 9. Consequently, for $k_\perp \rho<1$, contributions from modes with $l>1$ are strongly suppressed, scaling at least as the sixth power of the small argument. Therefore, for this particular source, the validity of a Taylor expansion of Eq.~(\ref{eq:SecBessel}) in powers of $k_\perp \rho$ constrains the radial distance to the vortex axis to remain sufficiently small, that is $\rho < 1/k_\perp $. Namely,
\be
\frac{\rho}{\lambda}<\frac{1}{2\pi}\sqrt{1+\frac{z^2}{R^2}},
\ee
which reduces to $2\pi R\rho<z\lambda$ further away from the source (i.e., in a paraxial limit $z\gg R$). Comparing the above condition with that of far-field (Fraunhofer) diffraction from an aperture of radius $R$, $z\gg R^2/\lambda$, and recalling that $R\gg \lambda$, we conclude that nondiffractive skyrmionic textures of wavelength-scale size can form in the near-field region, close to the source and around the vortex axis, well before the far-field conditions are reached.

\section{Optical Bessel vortex beams}

As mentioned in the main text, in optics, there are differences for the spin vector field depending on the chosen modal basis. This is illustrated by considering optical Bessel vortex beams formed by superposition of plane waves, with helicity $\sigma$, whose
wavevectors lie on a conical surface making an angle $\theta$ with respect to the propagation axis $z$ \cite{Durnin87,Jaregui05},
\begin{align}
\label{eq:twistedwf}
& \vect{E}^{\rm Bessel}_{\kappa l\sigma}(\vect{r}) = i \sigma \, E_0 \,
	e^{i(k_z z + (l+\sigma) \phi )}				\Bigg\{	e^{-i \sigma \phi}  \cos^2\frac{\theta}{2} 	\,
	J_{l}(\kappa\rho) \, \vect{u}_\sigma 
				\nn\\
	&  - \frac{i}{\sqrt{2}}  \sin\theta	\,
	J_{l+\sigma}(\kappa\rho) \, \vect{u}_z 
	-   e^{ i \sigma \phi}  \sin^2\frac{\theta}{2} 	\,
	J_{l+2\sigma}(\kappa\rho) \, \vect{u}_{-\sigma}
	\Bigg\}	\,,
\end{align}
where $\kappa=k\sin \theta$, $k_z=k \cos\theta$, and $\vect{u}_\sigma = (\vect{u}_x+ i\sigma\vect{u}_y )/\sqrt2$. Using a paraxial approximation $\theta\ll 1$ and Taylor expansion for Bessel functions $J_n(x)\approx\frac{1}{n!}(\frac{x}{2})^n$, we find that for $l=1$ core expressions for the field are same  for Bessel and Laguerre-Gauss, but differ for $l>1$. Namely, in the core region, we have for $\sigma l <0$
\bea
\nonumber
\vect{E}^{\rm Bessel}_{\kappa \sigma}(\vect{r})&\propto e^{-i\sigma\phi} (k\rho)^2\vect{u}_{\sigma} - i|l|\sqrt{2}(k\rho)\vect{u}_{z}\\
&- |l|(|l|-1)e^{i \sigma\phi}\vect{u}_{-\sigma}\,,
\eea
which leads to the following core expression for the unit spin vector field,
\be
\label{eq:BesSpinCore}
\vect{\widetilde s\,}^{\rm Bessel}_{\rm light} = \frac{[0,-2|l|\xi,\sigma(1-|l|(|l|-1)\xi^2)]}{\sqrt{1+2|l|(|l|+1)\xi^2+l^2(|l|-1)^2\xi^4}}\,.
\ee
This gives half-integer Skyrme number $N=-\sigma/2$ for $l=1$ and an integer Skyrme number $N=-\sigma$ for $|l|>1$ Bessel vortex. In contrast, Bessel solution for an acoustic vortex yields $N=\pm 1/2$ for any $|l|\geq1$.


\end{document}